# COSMOLOGICAL ADVENTURES IN THE LYMAN FOREST [*]

STEFANO CRISTIANI

Dipartimento di Astronomia Università di Padova
and
European Southern Observatory

## 1   Introduction

The observation and interpretation of absorption lines in QSO spectra has become a major topic of modern cosmology. QSO absorption lines, in fact, allow to probe the Universe, along the lines of sight to the most distant and luminous objects known, up to redshifts close to 5, when it was only 7 percent of its present age and one-sixth of its present size. This represents a unique and independent source of information, between the cosmic microwave background (CMB) data, at $z \sim 1000$, and the large scale distribution of galaxies, at $z \lesssim 0.5$, a Pandora-box containing answers to questions like:
What were the physical conditions of the primordial universe ?
What fraction of the matter was in a diffuse medium and what fraction and how early condensed in clouds ?
What fraction and types of dark matter were there ?
When and how the formation of galaxies and of large scale structure started ?
How much metals were produced and how early ?
What was the typical radiation field, how homogeneous, what was producing it ?
Does the standard big bang model make the correct predictions about primordial element abundances and CMB evolution ?
How dust obscuration and gravitational lensing affect our view of the distant Universe?

   The story of QSO absorption lines started in 1965 when Gunn & Peterson [1] and Bahcall & Salpeter [2] suggested that QSOs could be used to study the intervening intergalactic medium. In particular the presence of diffuse neutral hydrogen was expected to be revealed by an absorption trough bluewards of the Lyman-$\alpha$ emission.

   One year after, Burbidge and collaborators [3] detected the first absorption lines in the spectrum of the QSO 3C 191, and a controversy started to decide whether such lines were due to gas emitted by the QSO itself or originated by intervening material. In 1971 Lynds [4] obtained a spectrum of 4C 05.34, one of the first QSOs found with a redshift larger than 2.5. At this relatively high redshift the region bluewards of the Lyman-$\alpha$ emission becomes accessible to ground observations and in this range Lynds discovered what he called a "forest" of absorption lines, much more numerous than in the region longward the Lyman-$\alpha$ emission. He correctly attributed these lines to a large number of intervening Lyman-$\alpha$ absorbers.

---

[*]Lecture presented at the International School of Physics "Enrico Fermi" Course "Dark Matter in the Universe", Varenna, 25 July - 4 August, 1995.

In the following decade considerable progresses where made and a classification was elaborated that has basically survived till the present days [5], gradually shifting the focus from the dispute on the intrinsic-extrinsic nature of the absorbers to their cosmological properties.
According to this classification three types of absorbers have been defined:

1) the broad absorption lines, thought to be intrinsic, emitted by the QSO, by analogy with P-Cyg profiles in stars.

2) the sharp metal absorption lines, typically identified on the basis of MgII and CIV doublets, in part associated with the QSO when their redshift is close to the emission redshift, and in most cases intervening, due to galaxies intercepted along the line-of-sight.

3) the Lyman-$\alpha$ forest, ascribed to a sort of primordial intergalactic clouds either gravitationally or pressure confined by an hot IGM.

The two basic arguments on which the metal systems - Lyman-$\alpha$ systems dichotomy rests are: the absence of metal absorptions in correspondence of Lyman-$\alpha$ systems and the absence of clustering of Lyman-$\alpha$ lines, in contrast with the metal absorbers that have been observed to cluster on scales up to 600 km/s [40]. It should be noted that metal systems do show the corresponding Lyman-$\alpha$ absorption. When a galaxy is intercepted in a way that a typical metal system is observed, it is expected to give also origin to a strong Lyman-$\alpha$ absorption, with column density often larger than $10^{17}$ cm$^{-2}$ and therefore optically thick at the Lyman limit, with a characteristic edge. In rarer and more spectacular cases, when the disk of a galaxy is encountered, damped Lyman-$\alpha$ absorptions (DLAS) are observed, which derive their name from the damped wings of a Voigt profile becoming evident at column densities of the order $10^{20}$ cm$^{-2}$. In this way the Lyman limit and the damped Lyman-$\alpha$ systems have been considered to belong to the class of the metal systems.

The special interest of observations of Lyman-$\alpha$ absorptions lies in their high sensitivity: with the present instrumentation it is possible to detect neutral HI column densities down to $10^{12}$ cm$^{-2}$, while for example 21 cm radio observations are limited in the best cases to column densities that are 7 order of magnitude larger. In this way observations of the Lyman-$\alpha$ forest reveal very different structures, ranging from fluctuations of the diffuse intergalactic medium to the interstellar medium in protogalactic disks.

## 2  A Lyman-$\alpha$ database

The typical instrumentation for observations of Lyman-$\alpha$ absorptions consists of big telescopes and big echelle spectrographs, working at resolutions of the order $R \gtrsim 20000$. Most of the data used in the following discussion come from an ESO key-programme devoted to the study of QSO absorption systems at high redshifts, which has been carried out at the New Technology Telescope with the ESO Multi-Mode Instrument.

Up to now spectra of six QSOs, at resolutions between 9 and 14 km s$^{-1}$, have been reduced (see Fig. 1). All the spectra have been analyzed in a uniform way and all the lines have been fitted using the FITLYMAN program which is now available in the ESO-MIDAS package [6]. It performs a $\chi^2$ minimization to derive the redshift $z$, the Doppler parameter $b$ † and the column density $N$ for isolated lines and individual components of the blends. Whenever possible, the Lyman-$\beta$ lines have been used to constrain the number of components in the strong saturated Lyman-$\alpha$ blends.

The S/N is typically greater than 10 per pixel element (corresponding to 15 per resolution element) and raises up to S/N∼ 20 − 60 per resolution element in the regions near the QSO Lyman-$\alpha$ emission.

We have complemented our data with other spectra available in the literature with similar resolution, S/N and redshift range, obtaining a final sample of 10 QSOs totaling more than 1100 Lyman-$\alpha$ lines [7] that in the following will be referred to as the *extended sample*.

## 3  The $b - N_{HI}$ distributions: the temperature of the Lyman-$\alpha$ clouds

We want to study the physical properties of the clouds. The first thing one can do is to check if the observed distributions of the line parameters are compatible with the standard picture of the Lyman-$\alpha$ absorbers as warm, extended, highly ionized gas clouds [8]. For example plotting Doppler parameters versus column densities.

A diagram of this type is historically important because in the past has been used to challenge the standard model. The existence of a correlation in the $b - N_{HI}$ plane would imply that the width of the lines is largely due to macroscopic motions, and therefore the clouds are colder than what the Voigt profile fitting seems to show. Besides, the simple existence of a significant fraction of lines with Doppler $b$ values less than 18 km/s, corresponding to temperatures $T \lesssim 20000$ K, would tell us that the clouds are relatively cold.

Cold clouds would be difficult to reconcile [9,10] in the framework of the standard model [8] with the large cloud sizes, $R \sim 100 - 200 h_{50}^{-1}$ kpc, inferred from the observations of QSO pairs or gravitational lenses [11,12,13,14]. In the past there have been claims in this sense [9], leading to the introduction of particular cooling mechanisms [15] or time dependent photoionization codes [16].

In Fig. 2 the $b - N_{HI}$ distributions derived from one of the QSOs of the ESO KP, 0000-26, is shown. From these data no evidence is derived for a $b - N_{HI}$ correlation. The zone of avoidance in the upper left part of the diagram is due to a selection effect: shallow, low-column density lines are hidden in the noise. There is only an indication

---

†The so-called *Doppler parameter b* corresponds, in the case of pure thermal broadening of the lines, to $\sqrt{2KT/m}$, and is usually expressed in km/s. To give an idea of the numbers, $b^2$ is equal, for HI lines, to $1.65 \cdot 10^{-2}\ T$.

of a lack of high-column density, low-b lines, that would be easy to detect.

Fig. 3 shows the histograms of the Doppler parameter for Q0000-26 and PKS2126-158. The bulk of the Doppler parameters lie in the range $b = 20 \div 30$ km s$^{-1}$ with a fraction of about 15% of the lines with $b < 20$ km s$^{-1}$. Again, the present data are not inconsistent with the standard model: the low-b wing of the distribution can be ascribed to the effects of measurement errors, the very few cases ($\lesssim 2\%$) of lines with $b < 10$ km s$^{-1}$, may be representative of unrecognized metal lines.

On the other hand the existence of large thermal Doppler parameters is inconsistent with the standard model: large, hot clouds would be very highly ionized and summing over all the clouds the observed neutral and the ionized matter would exceed the baryon limit from nucleosynthesis $\Omega_b < 0.04 h_{50}^{-2}$. This argument places an upper limit to the typical Doppler parameter of the order $b < 30$ km/s and suggests that the large-b lines are blends of unresolved lower-b lines [17].

## 4 The HI column density distribution

The column density distribution of the lines in the *extended sample* is shown in Fig. 4. It can be represented by a double power-law with a break at $\log N_{HI} \sim 14$. At lower column densities the slope is flat, with an index $\beta_f \sim 1.1$, and above the break becomes steeper with $\beta_s \sim 1.8$. There are two main biases that do alter the shape of the column density distribution. The strong saturated lines above the break often show complex structures and to be deblended require information on the corresponding Lyman-$\beta$ lines, which are often outside the available spectral range or contaminated by lower-$z$ Lyman-$\alpha$ lines. The weak lines below the break are affected by the so-called "line-blanketing": high column density lines conceal weak lines. This latter bias becomes more and more important at higher redshifts.

Once the blanketing effect is quantified, via simulations, which we have extensively carried out with the FITLYMAN code, it is possible to correct the column density distribution, which becomes as shown in Fig. 4 by the dashed points. The slope below the break is increased to $\beta_f \sim 1.4$, a value consistent with previous estimates obtained at lower redshifts [18] where line blanketing effects are less important.

But the distribution is still inconsistent with a single power-law. The effect is not due to an underestimate of the blanketing but rather to a deficiency of lines at larger $N_{HI}$, as becomes apparent by comparing the present data with the data compiled by Petitjean et al. [19] at much larger column densities. There is a significant deficit of clouds in the range $10^{14} - 10^{17}$ cm$^{-2}$ and not only the power-law slope becomes again flat at larger column densities, but an overall representation with a power-law ($\beta \sim -1.5$) is remarkably good (except for the above mentioned deficit) over nearly ten orders of magnitude in column density [17].

## 5 The Redshift Distribution

The observed number density of Lyman-$\alpha$ clouds with $\log N_{HI} \gtrsim 14$, i.e. beyond the break of the column density distribution, is plotted in Fig 5a as a function of the redshift, together with that of the damped systems [20]. Assuming a standard power-law evolution of the type $dn/dz \propto (1+z)^\gamma$, Storrie-Lombardi et al. [20] found $\gamma = 1.5$ for the DLAS, while the slope for the Lyman-$\alpha$ clouds in the *extended sample* ($z \gtrsim 2$) corresponds to $\gamma = 2.7$. Besides, the normalizations of the number densities differ by about two orders of magnitude.

We can use the information about the sizes of Lyman-$\alpha$ clouds and DLAS to estimate the corresponding comoving volume densities. Bechtold et al. [21] give, assuming spherical geometry, a median value for the radius of Lyman-$\alpha$ clouds of the order of 200 kpc. For the DLAS, typical HI column densities $\geq 2 \times 10^{20}$ cm$^{-2}$ imply disks of $\lesssim 25$ kpc radii [22], as also confirmed by direct imaging [23,24].

With such sizes and geometries the volume densities of the Lyman-$\alpha$ clouds with $\log N_{HI} \gtrsim 14$ and of the DLAS appear remarkably similar, as shown in Fig. 5b. Also their cosmological evolutions, if we add to the *extended sample* points the data produced by the HST KP on QSO absorption lines [25] appear consistent.

Thus, it is tempting to associate Lyman-$\alpha$ clouds with $N_{HI} \gtrsim 10^{14}$ cm$^{-2}$ to the outer regions of the galaxies responsible for the damped systems, as also suggested by low-redshift data [26].

## 6 The measure of the ultraviolet background from the proximity effect

The redshift evolution of the line number density, within an individual spectrum, does not follow the general cosmological trend when approaching the QSO emission redshift: the density of lines, instead of increasing, decreases. This "inverse effect" has been interpreted and modeled as a "proximity effect" [27,28], i.e. the reduction of the line density in the region near the QSO emission is ascribed to the increase of the ionizing flux due to the QSO itself. In this way, absorbers near QSOs are more highly ionized than those farther away, where the general ultraviolet background (UVB) is the only source of ionization. The proximity effect is thus a powerful method to measure the intensity of the metagalactic ionizing radiation at high redshifts.

We can take the standard parameterization of the line distribution as a function of the redshift and column density

$$\frac{\partial^2 n}{\partial z \partial N_{HI}} = A_o (1+z)^\gamma \begin{cases} N_{HI}^{-\beta_f} & N_{HI} < N_{break} \\ N_{HI}^{-\beta_s} N_{break}^{\beta_s - \beta_f} & N_{HI} \geq N_{break} \end{cases} \qquad (1)$$

and generalize it, including the proximity effect. Near the QSO, highly ionized clouds are observed with a column density

$$N_{HI} = \frac{N_\infty}{1+\omega} \quad (2)$$

i.e. the intrinsic column density $N_\infty$, which the same cloud would have at infinite distance from the QSO, divided by the factor $1 + \omega$, where:

$$\omega(z) = \frac{F}{4\pi J} \quad (3)$$

is the ratio between the flux $F$ that the cloud receives from the QSO and the flux $J$ that the cloud receives from the general UVB [27].

Considering the conservation law for the number of lines

$$f(N) = g(N_\infty)dN_\infty/dN = g(N_\infty)(1+\omega) \quad (4)$$

one can write a general expression for the distribution of the lines and obtain a simultaneous estimate of the Lyman-$\alpha$ parameters of the $N_{HI}$, $z$ distributions *and* of the UVB $J_{LL}(z)$ evaluated at the Lyman limit.

$$\frac{\partial^2 n}{\partial z \partial N_{HI}} = A_o(1+z)^\gamma (1+\omega)^{1-\beta_f} \begin{cases} N_{HI}^{-\beta_f} & N_{HI} < N_{break} \\ N_{HI}^{-\beta_s} N_{break}^{\beta_s - \beta_f} & N_{HI} \geq N_{break} \end{cases} \quad (5)$$

where

$$N_{break}(z) = \frac{N_{\infty,b}}{1+\omega(z)}$$

is the observed break, which is shifted to lower and lower $N_{HI}$ as the QSO emission redshift is approached,

In this way a new parameter is introduced: the specific intensity $J$ of the UVB flux at 13.6 eV, which is contained in the variable $\omega$.

The measure of $J$ depends critically on the estimates of the flat slope of the column-density distribution, $\beta_f$, and on the cosmological evolution rate, $\gamma$. For this reason, it is not correct to separate the analysis of the Lyman-$\alpha$ line distribution from the estimate of $J$, as done in previous analyses, and a global Maximum Likelihood Analysis has to be carried out.

Table 1 reports the results of such a ML analysis. The first line shows the relevant parameters estimated without correcting for the blanketing effect. In the second row we find the results once the blanketing correction is introduced, and the variations give an idea of the importance of carrying out a global ML analysis. The blanketing correction of course steepens the column density distribution below the break, as we saw in Section 4, but also the rate of the redshift evolution is significantly increased. Because of these changes, the intrinsic number density of lines far away from the QSOs increases, especially at high redshifts, and this in turn causes a decrement of the value of the UVB derived from the proximity effect, that becomes $J_{-22} = 5 \pm 1$.

This value of $J$ is a factor of 6 lower than previous estimates and is remarkably close to the one predicted for the integrated contribution of QSOs [29].

We have repeated the ML analysis allowing for a power-law redshift evolution of the UVB of the kind

$$J = J_{(z=3)} \left(\frac{1+z}{4}\right)^j \quad (6)$$

The result is consistent with no evolution in the redshift interval $z = 2 - 4$, with still large uncertainties.

Using the $5 \pm 1$ value of $J_{22}$ together with the $\tau < 0.05$ $1\sigma$ upper limit for the GP optical depth at $z \sim 4.5$, it is possible to derive a value for the density of the diffuse part of the intergalactic medium, which turns out to be rather low, $\Omega_{IGM} \lesssim 0.01$, suggesting that most of the baryons are already in bound systems at $z \sim 4$ [30].

## 7 Clustering properties of the Lyman-$\alpha$ clouds

Systematic studies of the clustering properties of the QSO Lyman-$\alpha$ forest began in the early 1980's with the work by Sargent et al. [8], which concluded that no departures from random distributions of redshifts could be identified. Almost all the subsequent investigations have failed to detect any significant correlation on velocity scales $300 < \Delta v < 30000$ km s$^{-1}$ [31,32,33]. On smaller scales ($\Delta v = 50 - 300$ km s$^{-1}$) there have been indications of weak clustering [34,35,36,37], but this result appears controversial and relevant non-detections of clustering have also been reported [9,39]. On the contrary, metal-line systems selected by means of the CIV doublet [40] have been early recognized to show clustering on scales up to 600 km s$^{-1}$, suggesting a different spatial distribution.

To study the clustering we have adopted the two-point correlation function, defined as the excess, due to clustering, of the probability $dP$ of finding a Lyman$-\alpha$ cloud in a volume $dV$ at a distance $r$ from another cloud

$$dP = \Phi_{Ly\alpha}(z)dV[1 + \xi(r)] \quad (7)$$

where $\Phi(z)$ is the average space density of the clouds as a function of $z$.

In practice observations provide the redshifts of the Lyman-$\alpha$ lines that, due to peculiar motions, are not immediately transformed in distances. Therefore it is normally preferred to compute the two-point correlation function in the velocity space, estimated as

$$\xi(\Delta v) = \frac{N_{obs}(\Delta v)}{N_{exp}(\Delta v)} - 1 \quad (8)$$

where $N_{obs}$ is the observed number of line pairs with velocity separations between $\Delta v$ and $\Delta v + \epsilon_v$ and $N_{exp}$ is the number of pairs expected from a random distribution in redshift.

In our case $N_{exp}$ is obtained averaging 1000 numerical simulations of the observed number of lines, trying to account for all the relevant cosmological and observational effects.

The resulting correlation function for all the Lyman−$\alpha$ lines is shown in Fig. 6a. A weak but significant signal is present with $\xi \simeq 0.2$ in the 100 km s$^{-1}$ bin: 739 pairs are observed while 624 are expected for a random distribution, a 4.6$\sigma$ deviation from poissonianity. We have explored the variations of the clustering as a function of the column density. In Fig. 6b the correlation function for lines with $\log N_{HI} \leq 13.6$ is shown. All the evidence for clustering has disappeared. On the contrary, for lines with $\log N_{HI} \geq 13.8$ (Fig. 7), the correlation function at $\Delta v = 100$ km s$^{-1}$ shows a remarkable increase in amplitude ($\xi \simeq 0.6$) and significance: 234 pairs are observed while only 145 are expected for a random distribution, a more than 7$\sigma$ deviation from poissonianity. Fig. 8 shows the variation of the amplitude of the two-point correlation as a function of the column density threshold for the sample of the Lyman-$\alpha$ lines. On the same plot, for comparison, the two-point correlation function derived for CIV metal systems from the work by Petitjean and Bergeron [41] is reported. An extrapolation of the increasing amplitude trend observed for the two-point correlation function of the Lyman-$\alpha$ lines would easily intercept the corresponding estimate derived from the CIV metal systems. It is also interesting to note the similarity between the shapes of the correlation functions of Lyman-$\alpha$ and CIV systems, when observed at comparable resolution (cfr. Fig. 4b of [41]); and also the fact that a trend of increasing correlation with increasing column density is observed also in CIV systems, for which the strong systems appear usually in clumps of many components, while the weak doublets (generally optically thin at the Lyman limit) are typically made of an isolated, single component.

It is also possible to address the evolution of the two-point correlation function with redshift for Lyman-$\alpha$ lines with column densities $\log(N_{HI}) > 13.8$. The amplitude of the correlation at 100 km s$^{-1}$ decreases with increasing redshift from $0.85 \pm 0.14$ at $1.7 < z < 3.1$, to $0.74 \pm 0.14$ at $3.1 < z < 3.7$ and $0.21 \pm 0.14$ at $3.7 < z < 4.0$. Unfortunately, HST data are still at too low-resolution [25] or are too scanty [42] to allow a meaningful comparison with the present data.

Voids in the Lyman-$\alpha$ forest provide a test for models of the large-scale structure in the Universe and for the homogeneity of the UV ionizing flux. Previous searches for megaparsec-sized voids have produced a few claims [43,44,37], but great care has to be taken in the statistical approach to avoid the pitfalls of "a posteriori statistics". Detections of voids in individual cases are interesting but one would like to follow a more general approach, assessing more quantitatively how common is the phenomenon. Using the *extended sample* of Lyman-$\alpha$ lines it is possible to check in how many cases at least one void is observed, significant at least at the 5% level. This happens for 3 QSOs out of 10 (in one case 2 voids are observed in the same spectrum) corresponding to a binomial probability of 0.01. It is apparent that the regions corresponding to the voids are not completely devoid of lines: weak absorptions are observed within the voids [37]. This agrees with low-redshift observations [45], showing that in the local Universe voids are not entirely devoid of matter. Even if underdense regions are statistically significant in the *extended sample*, their filling factor is rather low: the above defined voids cover only about 2% of the available line-of-sight path-length.

In summary, the evidences accumulated about the sizes, evolution, clustering of the Lyman-$\alpha$ clouds with $\log N_{HI} \gtrsim 14$, as well as the observations of metallicities of the order $10^{-2}$ [46,47] corresponding to such systems, suggest a continuity scenario with a physical association between the Lyman-$\alpha$ clouds with $\log N_{HI} \gtrsim 14$ and the halos of protogalactic systems.

## 8 Simulations

Starting from these empirical evidences, it is becoming possible the comparison with another field that is reaching a remarkable predictive power: the N-body simulations.

Various codes have already been used in this way: tree smoothed-particle hydrodynamic [48], adapted particle mesh codes [49,50], total variation diminishing [51], the HERCULES code [52] etc.

The idea is simple: the simulation is run till a given redshift, random lines-of-sight are drawn within the data cube and the physical structures intercepted are examined, to see how well the observed distributions are reproduced. The model is further evolved to check the redshift evolutions.

A number of interesting results have already been produced. The observed column-density distribution is reproduced remarkably well, for example in a flat CDM model [53]. The break in the distribution may represent the transition from pressure to gravitational confinement [54], the passage from a variety of systems in various stages of gravitational infall and collapse or even from underdensities to gas associated with star forming galaxies [52]. This may help us understanding the similarities between damped, metal systems and Lyman-$\alpha$ lines with $N_{HI} > 10^{14}$ cm$^{-2}$. However, the same good agreement is obtained with a number of different models, and this generalized success suggests that the column density alone may be a rather insensitive test.

The distribution of Doppler parameters [53] and the clustering properties [49] are also qualitatively reproduced.

On the other hand all the models appear to be inconsistent with one or another of the constraints imposed by the observations of the Lyman forest (e.g. the intensity of the UV metagalactic flux) or with other more general constraints (e.g. the CDM models produce too much structure on scales of several 10's of Mpc, too many clusters [55]). Mixed dark matter models might be excluded by the observations of DLAS: they have difficulties in supplying enough collapsed structures to produce enough DLAS at $z > 3$ [50,56,57], a situation significantly exacerbated by reionization [58].

In the future we can expect quantities of developments and it is easy to predict that QSO absorption lines will play a crucial role in testing theories of cosmic structure formation.

# 9  Acknowledgements

I thank S. and V. D'Odorico, A. Fontana, E. Giallongo and S. Savaglio, members of the ESO KP collaboration, for valuable help. The partial support of the ASI contract 94-RS-107 is acknowledged.

**REFERENCES**

1. Gunn J.E., Peterson B.A., 1965, ApJ 142, 1633

2. Bahcall J.N., Salpeter E.E., 1965, ApJ 142, 1677

3. Burbidge 1966, ApJ 144, 447

4. Lynds R., 1971, ApJ 164, L73

5. Sargent W.L.W., 1988, in Blades C. J., Turnshek D. A., Norman C. A. eds. Proceedings of the *QSO Absorption Lines* Meeting, Baltimore 1987 May 19-21, Cambridge University Press, p.1

6. Fontana A., Ballester P., 1995, *The ESO Messenger* 80, 37

7. Giallongo E., Cristiani S., D'Odorico S., Fontana A., Savaglio S., 1995, ApJ submitted

8. Sargent W.L.W., Young P.J., Boksenberg A., Tytler D., 1980, ApJS 42, 41

9. Pettini M., Hunstead R. W., Smith L. J., Mar D. P., 1990, MNRAS 246, 545

10. Donahue M., Shull J.M., 1991, ApJ 383, 511

11. Smette A., Surdej J., Shaver P.A. et al., 1992, ApJ 389, 39

12. Smette A., Robertson J.G., Shaver P.A., Reimers D., Wisotzki L., Köhler Th., 1995, A&AS 113, 199

13. Bechtold J., Crotts A. P. S., Duncan R. C., Fang Y., 1994, ApJL, 437, L83

14. Dinshaw N., Foltz C.B., Impey C.D., Weymann R.J., Morris S.L. 1995, Nature 373, 223

15. Giallongo E., Petitjean P., 1994, ApJL, 426, L61

16. Ferrara A., Giallongo E., 1995 in preparation

17. Hu E. M., Kim T., Cowie L., Songaila A. Rauch M., 1995, AJ 110, 1526

18. Giallongo E., Cristiani S., Fontana A., Trevese D., 1993, ApJ 416, 137

19. Petitjean P., Webb J.K., Rauch M., Carswell R.F., Lanzetta K., 1993 MNRAS 262, 499

20. Storrie-Lombardi L.J., McMahon R.G., Irwin M.J., Hazard C., 1995, in G. Meylan ed. Proceedings of the ESO Workshop on *QSOs Absorption Lines*, Springer, p. 47.

21. Bechtold J., Crotts A. P. S., Duncan R. C., Fang Y., 1994, ApJL 437, L83

22. Broeils A.H., van Woerden H., 1994, A&AS 107, 129

23. Steidel C.C., Pettini M., Dickinson M., Persson S.E., 1994, AJ 108, 2046

24. Steidel C.C., Bowen D.V., Blades J.C., Dickinson M., 1995, ApJL, 440, L45

25. Bahcall J.N., Bergeron J., Boksenberg A. et al., 1995, babbage astro-ph/9506124

26. Lanzetta K. M., Bowen D. B., Tytler D., Webb J. K., 1995, ApJ, 442, 538

27. Bajtlik S., Duncan R. C., Ostriker, J. P., 1988, ApJ, 327, 570

28. Weymann R. J., Carswell R. F., Smith M. G., 1981, ARA&A, 19, 41

29. Haardt F., Madau P., 1995, babbage astro-ph/9509093

30. Giallongo E., D'Odorico S., Fontana A., McMahon R. G., Savaglio S., Cristiani S., Molaro P., Trevese D., 1994, ApJL, 425, L1

31. Sargent W. L. W., Young P. J., Schneider D. P., 1982, ApJ 256, 374

32. Bechtold J., 1987, in J. Bergeron et al. eds., Proc. Third IAP Workshop, *High Redshift and Primeval Galaxies*. Editions Frontieres, Gif-sur-Yvette, p. 397

33. Webb J. K., Barcons X., 1991, MNRAS 250, 270

34. Webb J. K., 1987, in Hewett A., Burbidge G., Fang L. Z. eds., Proc. IAU Symp. 124, *Observational Cosmology*. Reidel, Dordrecht, p. 803

35. Rauch M., Carswell R. F., Chaffee F. H., Foltz C. B., Webb J. K., Weymann R. J., Bechtold J., Green R. F. 1992, ApJ 390, 387

36. Chernomordik V.V., 1995, ApJ 440, 431

37. Cristiani S., D'Odorico S., Fontana A., Giallongo E., Savaglio S., 1995, MNRAS 273, 1016

39. Stengler-Larrea E. A., Webb J. K. 1993, in Chincarini G., Iovino A., Maccacaro T., Maccagni D. eds., *Observational Cosmology*, ASP Conference Series, 51, 591

40. Sargent W. L. W., Boksenberg A., Steidel C. C. 1988, ApJS 68, 539

41. Petitjean P., Bergeron J., 1994, A&A 283, 759

42. Brandt J. C., Heap S. R., Beaver E. A., et al., 1995, AJ 105, 831

43. Crotts A. P. S. 1989, ApJ 336, 550
44. Dobrzycki, A., Bechtold, J. 1991, ApJ 377, L69
45. Stocke J.T., Shull J.M., Penton S., Donahue M., Carilli C., 1995, ApJ 451, 24
46. Cowie L.L., Songaila A., Kim T., Hu E.M., 1995, AJ 109, 1522
47. Tytler D., Fan X.-M., 1995, ApJ submitted
48. Katz N., Weinberg D.H., Hernquist L., 1995, babbage astro-ph/9509107
49. Mücket J. P., Petitjean P., Kates R. E., Riediger R., 1995, babbage astro-ph/9508129
50. Klypin A., Borgani S., Holtzman J., Primack J., 1995, ApJ 444, 1
51. Cen R., Miralda-Escudé J., Ostriker J.P., Rauch M., 1994, ApJ 437, L9
52. Zhang Y., Anninos P., Norman M.L., 1995, babbage astro-ph/9508133
53. Hernquist L., Katz N., Weinberg D. H., Miralda-Escudé J., 1995, babbage astro-ph/9509105
54. Charlton J.C., Salpeter E.E., Linder S. M., 1994, ApJ 430, L29
55. White S.D.M., Efstathiou G., Frenk C., 1993, MNRAS 262, 1023
56. Mo H.J, Miralda-Escudé J., 1994, ApJ 430, L25
57. Ma C.P., Bertschinger E., 1994, ApJ 434, L5
58. Shapiro P. R., 1995 in McKee C.F., Ferrara A., Heiles C.E., Shapiro P.R. eds., *The Physics of the Interstellar Medium and the Intergalactic Medium*. ASP Conference Series, in press

Table 1. Maximum likelihood analysis for lines with $\log N_{HI} \geq 13.3$

| $N_l$ | $\gamma$ | $\beta_f$ | $\log N_{\infty,b}$ | $\beta_s$ | $\log J$ | $j$ |
|---|---|---|---|---|---|---|
| 1128 | 2.49±0.21 | 1.10±0.07 | 14.00±0.02 | 1.80±0.03 | -21.21±0.07 | |
| | 2.65±0.21 | 1.34±0.07 | 13.98±0.04 | 1.80±0.03 | -21.32±0.08 | – |
| | 2.67±0.26 | 1.33±0.09 | 13.96±0.07 | 1.80±0.04 | -21.32±0.10 | -0.28±1.41 |

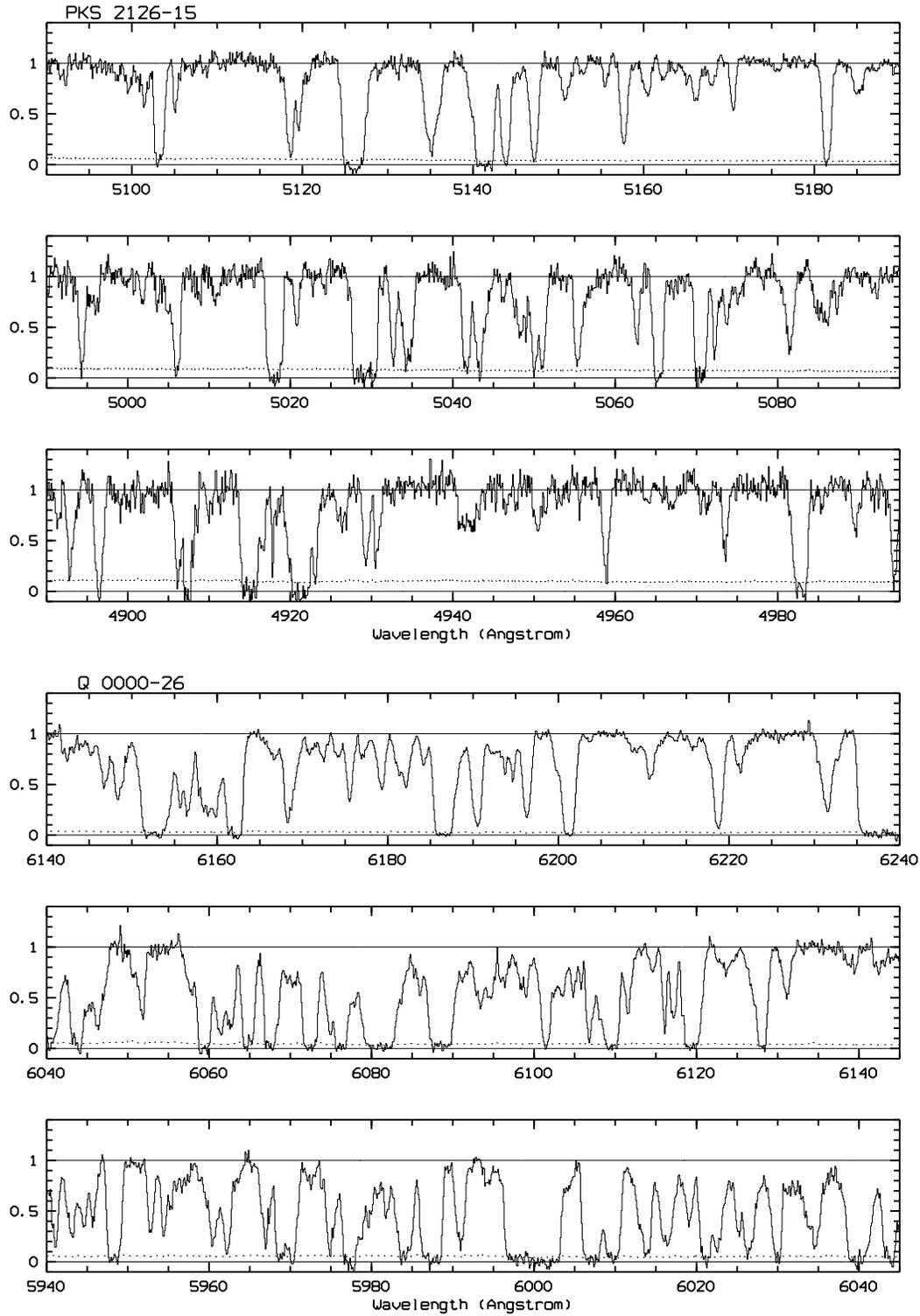

Figure 1: Normalized spectra of the regions near the Lyman-α emissions for PKS 2126-158 (upper plots) and Q0000-26 (lower plots). The dashed line shows the noise level per resolution element.

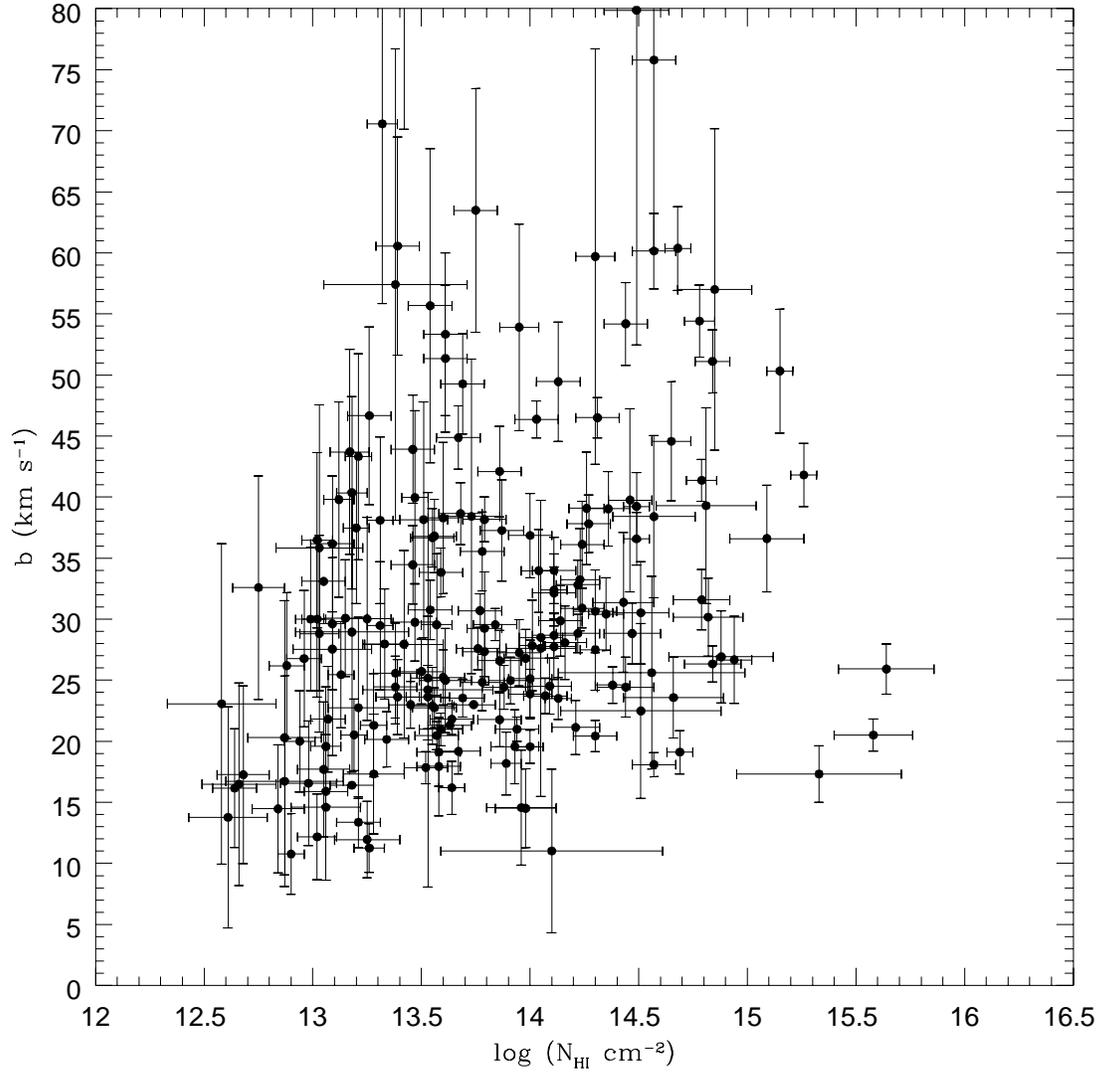

Figure 2: Doppler parameter - column density plane for Lyman-$\alpha$ lines at a distance larger than 8 Mpc from the QSO 0000-26.

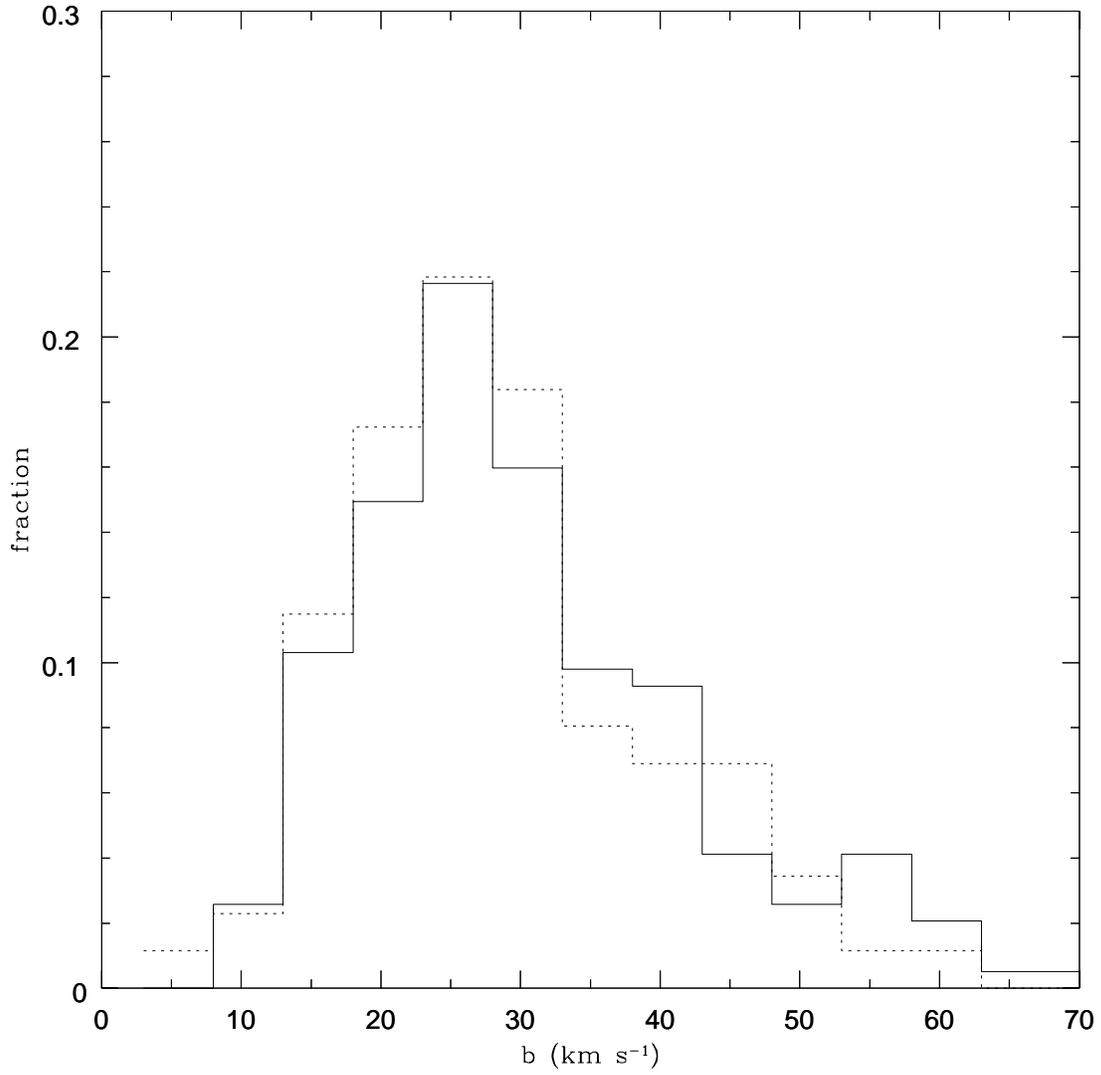

Figure 3: Doppler distributions of the lines of two QSOs of the ESO KP (dotted line PKS2126-158, continuous histogram Q0000-26).

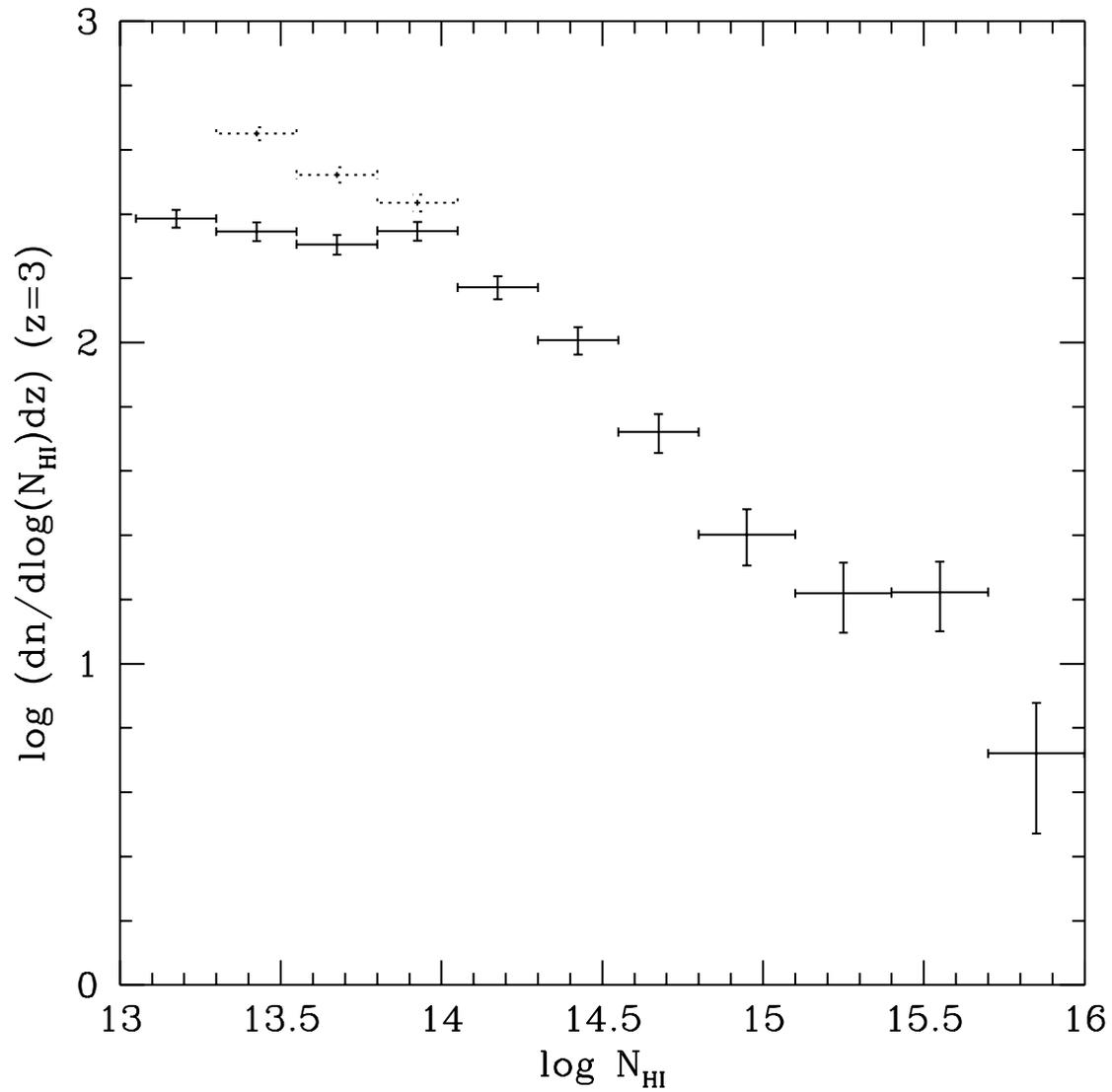

Figure 4: Column density distribution for the sample of Giallongo et al. [7] (continuous histogram). The dashed part shows the blanketing correction.

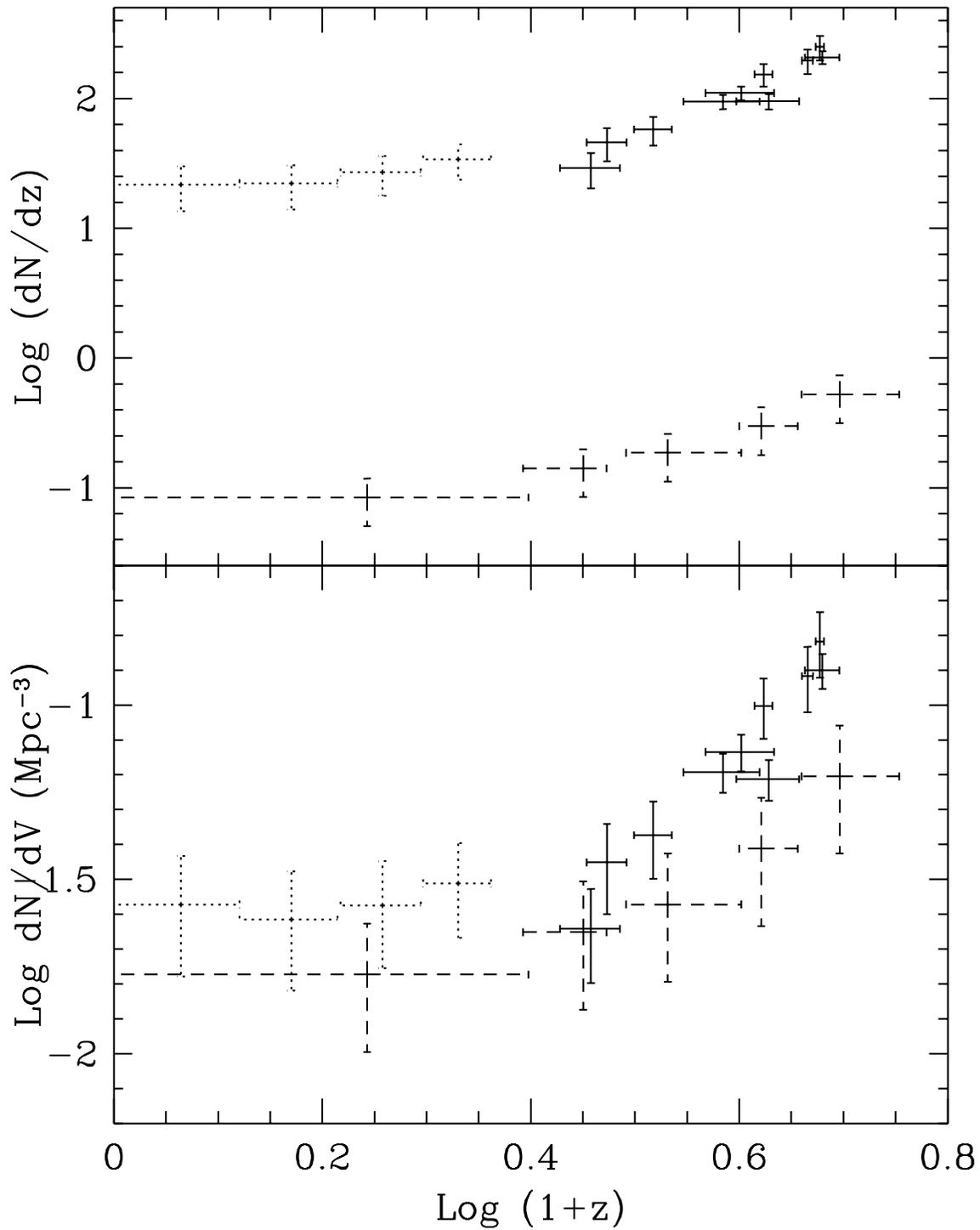

Figure 5: a) (upper panel) Redshift distribution of the Lyman-α lines with log $N_{HI} \geq 14$ [7,25] (upper points) together with the redshift distribution of the damped systems taken from Storrie-Lombardi et al. [20], (dashed lower points). b) (lower panel) Comoving volume density of the same samples as in a).

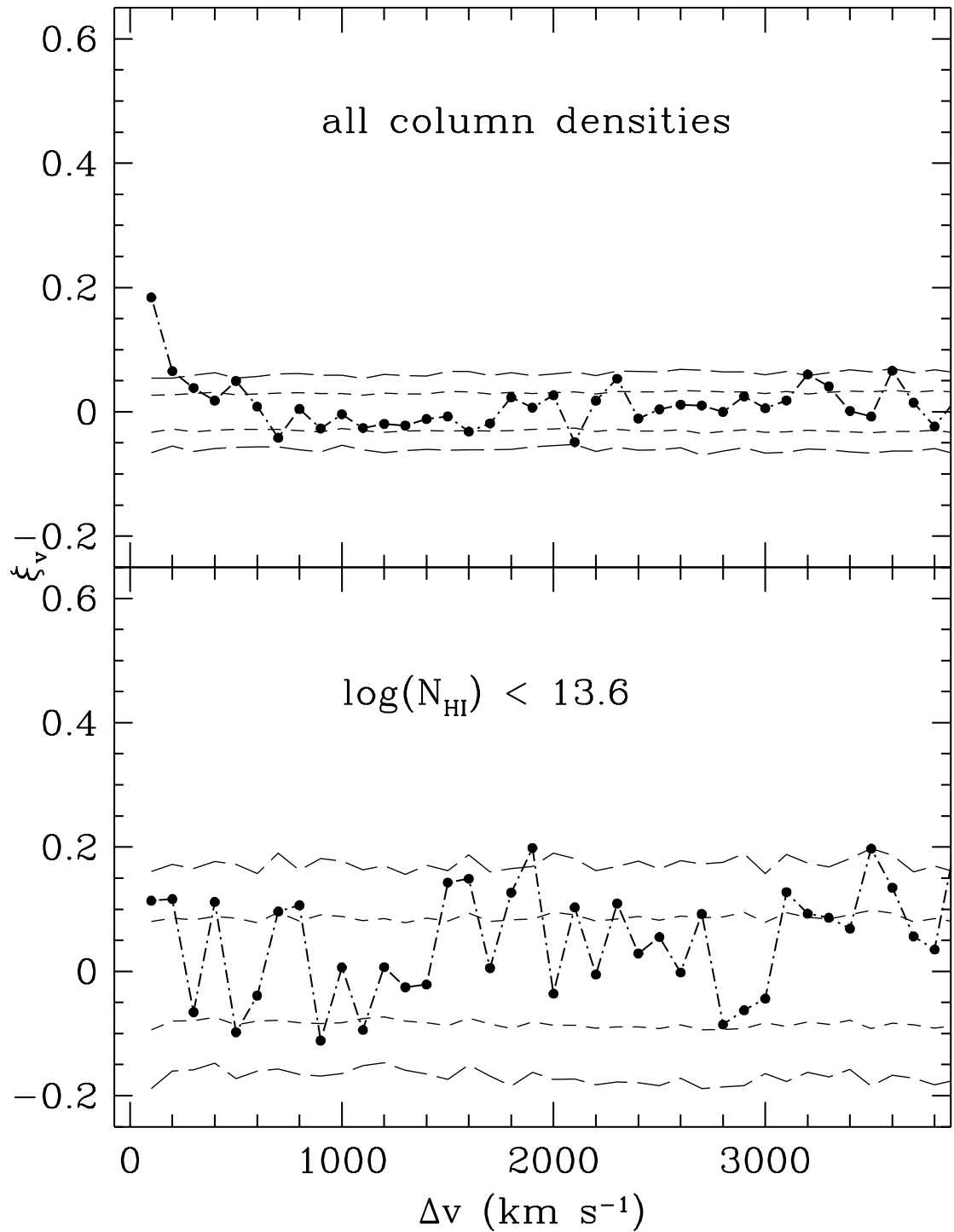

Figure 6: Two-point correlation function in the velocity space: *a)* (upper panel) for the complete sample of Lyman-$\alpha$ lines, *b)* (lower panel) for lines with column densities $< 10^{13.6}$ cm$^{-2}$. The short-dashed and long-dashed lines represent the $1\sigma$ and $2\sigma$ confidence limits for a poissonian process.

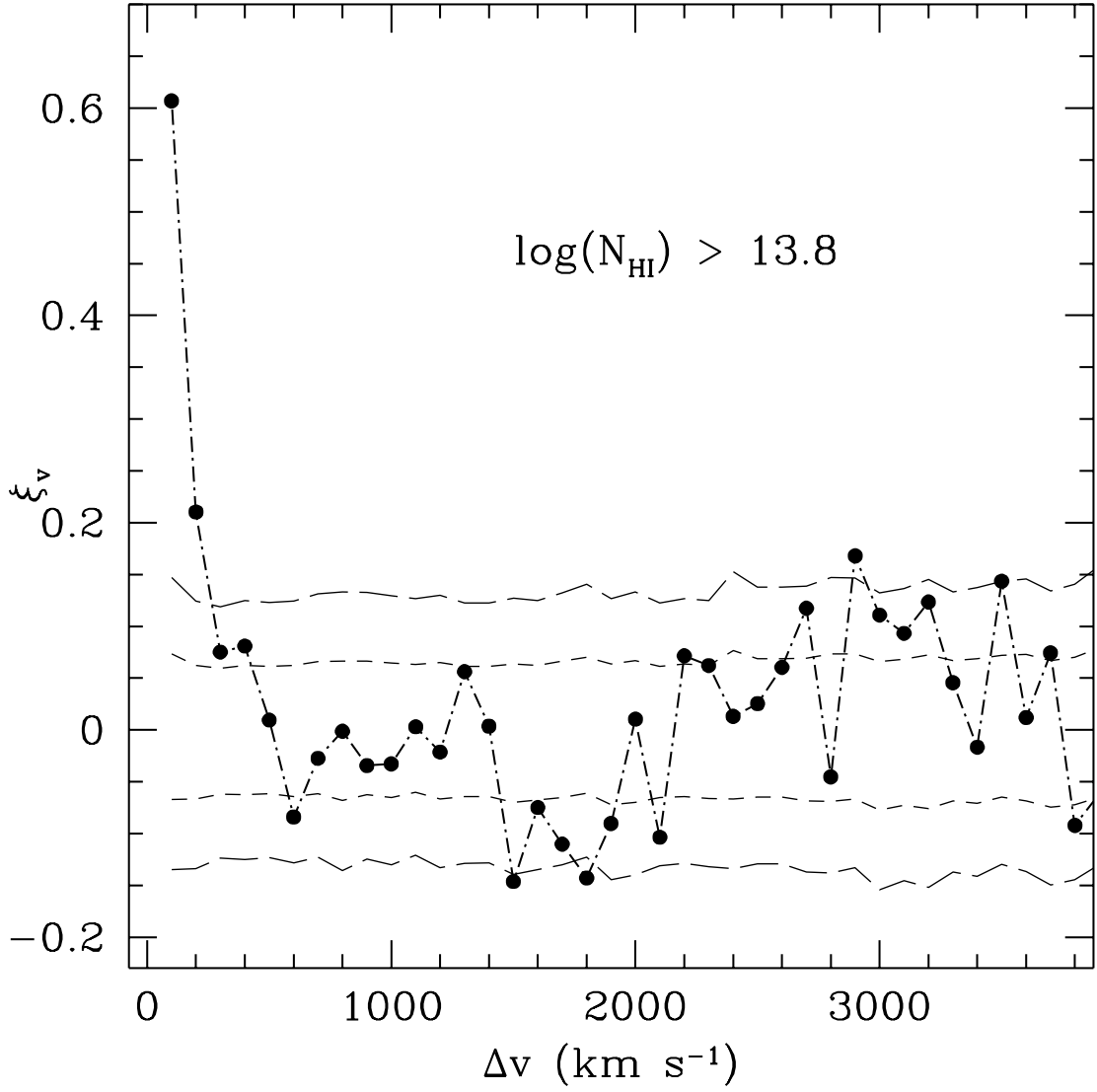

Figure 7: Two-point correlation function in the velocity space for lines with column densities $> 10^{13.8}$ cm$^{-2}$. Confidence limits as in Fig. 6

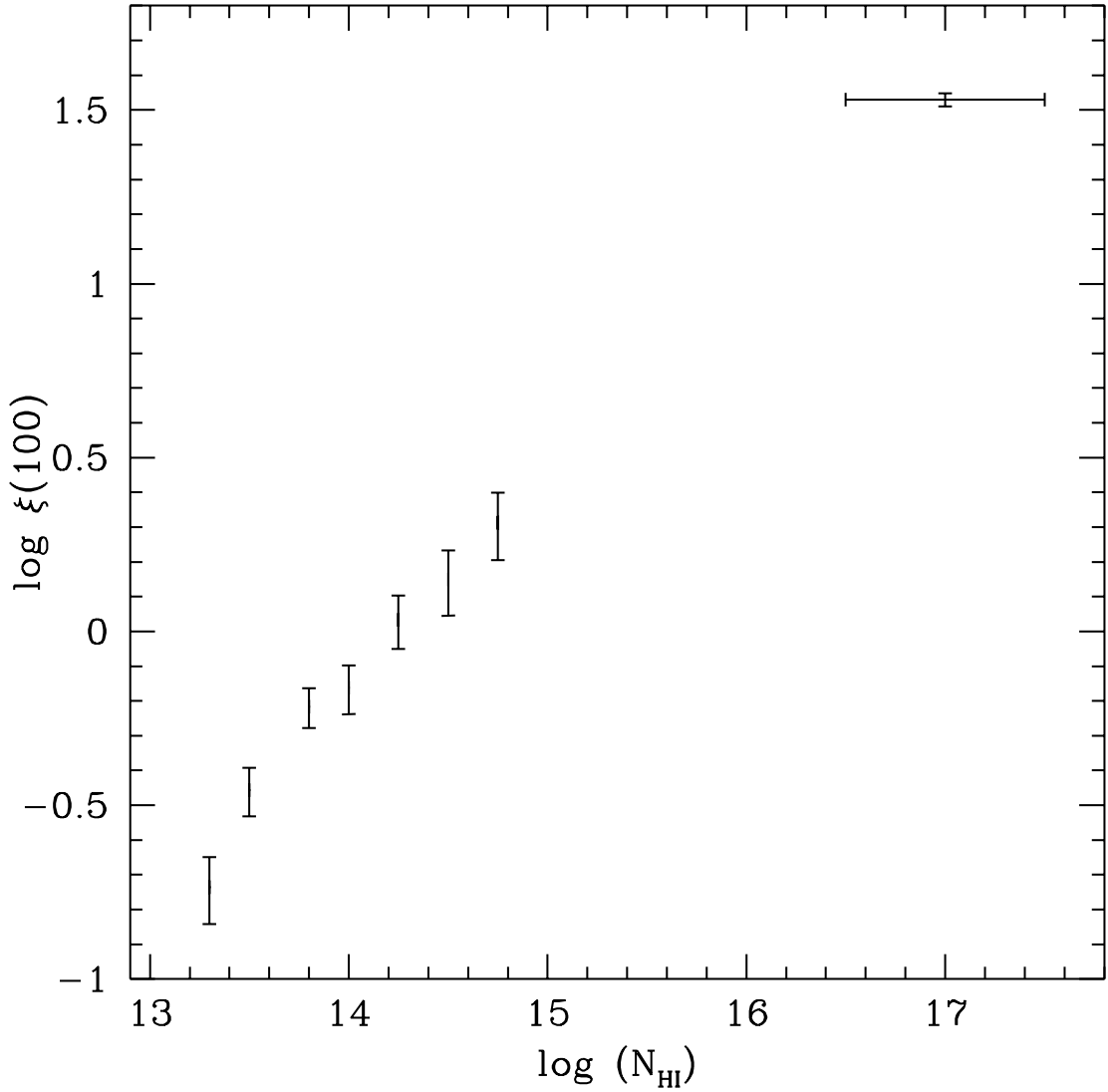

Figure 8: Variation of the amplitude of the two-point correlation function as a function of the column density threshold for the sample of the Lyman-$\alpha$ lines. The point on the upper-left side of the picture shows the correlation of the CIV metal systems.